# Optical and mechanical properties of amorphous Mg-Si-O-N thin films deposited by reactive magnetron sputtering

*Sharafat Ali*[a] *, *Biplab Paul*[b], *Roger Magnusson*[b], *Erik Ekström*[b], *Camille Pallier* [b]**, *Bo Jonson*[a], *Per Eklund*[b] *and Jens Birch*[b]

[a]*School of Engineering, Department of Built Environment and Energy Technology, Linnæus University, SE-351 95 Växjö, Sweden;*

[b]*Department of Physics, Chemistry and Biology, ( IFM), Linköping University, SE-58183 Linköping, Sweden*

* Corresponding Author: Dr. Sharafat Ali

Tel: +46-470-708991

Fax: +46-470-708756

E-mail: sharafat.ali@lnu.se

**Present address: RISE IVF, Bröderna Ugglas gata, 581 88 Linköping, Sweden.

**Abstract**

In this work, amorphous thin films in Mg-Si-O-N system typically containing > 15 at.%, Mg and 35 at.%, N were prepared in order to investigate especially the dependence of optical and mechanical properties on Mg composition. Reactive RF magnetron co-sputtering from magnesium and silicon targets were used for the deposition of Mg-Si-O-N thin films. Films were deposited on float glass, silica wafers and sapphire substrates in an Ar, $N_2$ and $O_2$ gas mixture. X-ray photoelectron spectroscopy, atomic force microscopy, scanning electron microscopy, spectroscopic ellipsometry, and nanoindentation were employed to characterize the composition, surface morphology, and properties of the films. The films consist of N and Mg contents up to 40 at.% and 28 at.%, respectively and have good adhesion to substrates and are chemically inert. The thickness and roughness of the films increased with increasing content of Mg. Both Hardness (16 – 21 GPa) and reduced elastic modulus (120 – 176 GPa) are strongly correlated with the amount of Mg content. The refractive index up to 2.01 and extinction coefficient up to 0.18 were found to increase with Mg content. The optical band gap (3.1 – 4.3) decreases with increasing the Mg content. Thin film deposited at substrate temperature of 100 °C shows a lower value of hardness ( 10 GPa), refractive index (1.75), and higher values of reduced elastic modulus (124 GPa) as compared to the thin film deposited at 310 °C and 510 °C respectively, under identical synthesis parameters.

## 1. Introduction

Reactive magnetron sputtering is an effective process to deposit oxides, nitrides and oxynitride thin films due to its advantages such as good uniformity, low deposition temperature and ability to deposit compact films with high mechanical/optical integrity and stability [1-4] For coatings on glass, oxynitrides are of particular interest [5-9]. Oxynitride coating is particular interest in optics, photonics, microelectronics, waveguide devices, anti-reflection coating, nonvolatile memories and dielectrics applications. Silicon oxynitride (SiON) and silicon aluminum oxynitride (SiAlON) thin films have been extensively studied, due to their mechanical, chemical, thermal, electronic, and optical properties. These properties typically vary greatly with the O/N substitution in most cases. In comparison with $SiO_2$, $Si_3N_4$ and Si(Al)ON systems, less research has been devoted to M-Si-O-N (M = Alkaline-earth or rare-earth cations) thin film synthesis [10, 11]. The addition of alkaline-earth or rare-earth cations to SiON thin films constitute a significant extension of silicon oxynitride compositions and also show high chemical versatility.

Bulk Mg-Si-(Al)-O-N glasses [12-15] are commonly prepared by melt quenching and exhibit superior mechanical, chemical, thermal and optical properties compared to their oxide counterpart glasses. However, as yet, the gains in properties are not worth to compensate for the increased efforts required to synthesized and process large quantities of bulk glasses. Furthermore, these glasses are difficult to synthesize with retained transparency in the visible region. The non-transparency of N rich bulk oxynitride glasses limits their use in the traditional optical glass applications, even though their hardness and elastic modulus are significantly improved. Recently, we have prepared nitrogen-rich thin films in the Mg-Si-O-N [10] and Ca-Si-O-N [11] systems, which are transparent in the visible region. They are also free from impurities such as elemental silicon particles and silicides, usually observed in nitrogen-rich bulk glasses. Metal-containing SiON thin films can be deposited on amorphous substrates as a protective coating.

We have previously reported the synthesis and initial property characterization of thin films in the Mg-Si-O-N system prepared by reactive RF magnetron sputtering in $Ar/N_2/O_2$ gas mixture (at $N_2$ flow rate of 20%). In the present work, we have prepared thin films in Mg-Si-O-N system at varied $N_2$ flow rate, substrate temperature and Mg/Si target ratio, to investigate the effects of these parameters on properties. The mechanical properties, e.g. hardness and reduced elastic modulus and the optical properties in term of refractive index, extinction coefficients and band gap are investigated.

## 2. Experimental Procedure

Mg-Si-O-N thin films were deposited on commercial soda-lime silicate float glass (air-side) as well as on reference silica wafers, and sapphire substrates by RF magnetron sputtering. The deposition system is described elsewhere [16, 17]. The thickness of the float glass substrate is 4 mm, while silica wafers and sapphire substrates have thickness of 1 mm and 0.5 mm, respectively. For experiments, individual pieces of 10 mm × 10 mm were prepared. Reactive sputter deposition from silicon (purity 99.99 %), magnesium (purity 99.95 %) was performed in an ultra-high vacuum (UHV) deposition system with a base pressure $<1\cdot10^{-5}$ Pa. For reactive sputtering, a mixture of oxygen ($O_2$), nitrogen ($N_2$), and argon (Ar) was used. The total gas flow was kept constant at 40 sccm, while the Ar and $N_2$ flow rate was varied. The $O_2$ flow rate (0.6 sccm) were kept constant in all series. Thin films in the Mg-Si-O-N system were prepared in four different sample series. The first series (M1) has Mg target power 80, 84 and 96 W, and $N_2$ flow of 20 % of the total gas flow [Ar+$N_2$+$O_2$]. The Si target of power 60 W was kept constant. The second series (M2) has Mg target power 60, 100 and 140 W and $N_2$ flow of 20 %. Some properties of the M2 series were previously reported [10], and are included in the present study for comparison. The third series (M3) has Mg target power 60, 100 and 140 W and $N_2$ flow of 30 %. All the samples in series M2 and M3 were prepared with constant Si target power of 100 W. M4 series were prepared with constant power of Si (50 W) and Mg (70 W) having substrate temperature of 100, 350 and 510 $^{o}$C. The Mg-Si-O-N thin films in all series were deposited during a time period of 2 hours. For reference purposes, thin films of SiN and SiON were also deposited on float glass, silicon wafers and sapphire substrates using identical conditions as used for series M2.

X-ray diffraction (XRD) θ/2 θ scans were performed to determine the amorphous/crystalline nature of the Mg-Si-O-N thin films using a Panalytical X'pert PRO MPD diffractometer equipped with a Cu Kα source operated at 40 kV and 40 mA. A light microscope (Olympus PMG3, Japan) equipped with a digital camera was used to observe the surface morphology of the samples. The microstructures of the samples were examined by back-scattered electron images using a JSM 7000F scanning electron microscope. The SEM was operated at acceleration voltages of 15 and 7 kV. Atomic force microscopy (AFM) was carried out with a Nanoscope 111, in contact mode. The AFM images were processed by use of a WSxM software. Topographic images were taken, at different magnifications and several images were taken over different sections of the sample. The surface composition of the thin films was investigated by X-ray photoelectron spectroscopy (XPS) technique on silica wafers substrate.

XPS analyses were performed with an Axis Ultra DLD instrument from Kratos Analytical (UK). The base pressure in the system during spectra acquisition was $1.1 \times 10^{-9}$ Torr ($1.5 \times 10^{-7}$ Pa). Monochromatic Al Kα radiation ($h\nu = 1486.6$ eV) was used and the anode power was set to 225 W. All spectra were collected from the area of $0.3 \times 0.7$ $mm^2$ area at the center of the sputter-cleaned region. The elemental concentrations were derived using a CasaXPS software employing Shirley-type background [18] and manufacturer's sensitivity factors.

The durability of the thin films was investigated by keeping some samples in moisture atmosphere at 85 °C for three days and reexamining by ellipsometry technique one year later after the first examination. The X'Pert Reflectivity software was used to determine the density of the samples by using a fitting model calculation on the experimental data, the density, thickness, roughness, and refractive index (δ and β) of both film and substrate were used as fitting parameters.

Hardness (H) and reduced elastic modulus ($E_r$), of the thin films were measured by nanoindentation using a Triboindenter Ti 950 instrument (Hysitron, Inc., Eden Prairie, MN, USA) on float glass substrate. A Berkovich diamond tip with total included angle of 142.3° and a half angle of 65.35° was used, with the indentation load of 1 mN, and penetration depth not exceeding 10% of the film thickness for films having thicker than 200 nm. The Berkovich diamond tip was calibrated on a fused-silica sample. Each sample was measured twelve times to get a statistically valid average value. The indentation procedure consisted of three steps: 1) loading to P max during 5 s, 2) hold for 2 s, and 3) unloading during 5 s. The average hardness and reduced elastic modulus with standard deviation were calculated by the method of Oliver and Pharr using the unloading elastic part of the load-displacement curve [19].

Mueller matrix spectroscopic ellipsometry (MMSE) was used to evaluate the refractive index ($n_r$), extinction coefficient (k), and band gap (E) of the Mg-Si-O-N thin films on float glass substrate. The measurements were done by using a Mueller matrix ellipsometer, the RC2®, from J.A. Woollam Co., Inc. The measurements were done in the wavelength range of 210–1690 nm and angles of incidence of 45°, 55°, 65° and 70° in reflection. The ellipsometry data analysis was performed with the software CompleteEASE, version 4.72, also from J.A. Woollam Co., Inc. and fitted with a Tauc-Lorentz model [20] for amorphous films to assess their optical properties.

## 3. Results and discussion

*3.1. Thin film stoichiometry and morphology.*

In this section and section 3.2, series M1, M2 and M3 are presented and discussed. The elemental concentration on the surface of the samples (determined by XPS) of Mg, Si, O, N in atomic percentage (at.%) are summarized in Table 1. Mg contents up to 28 at.% and N contents up to 40 at.% were incorporated into the Mg-Si-O-N thin films.

The Mg concentration increases with increasing Mg target power, as shown in Fig. 1. A similar trend was also observed in the Mg-Si-O-N [10] thin films prepared with $N_2$ flow of 20 %. All the obtained films in the Mg-Si-O-N system have N content > 30 at.%, thus being classified as nitrogen rich films based on the definition as $\geq$ 10 at.% for bulk oxynitride glasses. No systematic relation between N and Mg concentrations is observed for Mg content higher than 20 at.% in the films. This observation can be attributed to the formation of more Mg-O than Mg-N when the Mg concentration is higher in the films and agrees with those of previous studies of metal-containing bulk silicon oxynitride glasses that the N content initially increases with increasing the modifier content in the glass network [21-23]. Thin films in the M2 series have a higher N and lower Mg contents than the M1 and M3 series. The thin films in Mg-Si-O-N system have higher content of N and Mg as compared to the bulk glass prepared by melt quenching or sol-gel techniques in the same system [12, 24]. The high nitrogen content in the Mg-Si-O-N thin films as compared to the Ca-Si-O-N thin films [11] prepared under similar conditions might be due to the high affinity of Mg towards nitrogen. A similar trend has been observed in bulk Mg/Ca-Si-O-N glasses [12].

Figure 2a shows that the Mg-Si-O-N thin films are homogeneous, free from macroscopic impurities and are optically transparent in the visible region as compared with the bulk glass in the Mg-Si-O-N system Figure 2b. Generally, bulk oxynitride glasses in the Mg-Si-(Al)-O-N system, contains macroscopic impurities in the form of metallic silicide and/or free silicon particles dispersed in the glass network, and are translucent in the visible spectrum [12].

X-ray diffraction confirmed that all films deposited on float glass, silica wafers and sapphire substrates are X-ray amorphous. The surface structures of the films were investigated by using SEM, AFM and optical microscopy. The SEM cross-section morphology (Fig. 3a) and SEM back scattered electron image (Fig. 3b) of a 480 nm-thick film deposited with a Mg target power of 100 W presents a uniform and featureless structure without any visible defects, typical for amorphous films. Similar morphology was observed for all the studied films. The surface topography of all films, analyzed by AFM (Fig. 3c), is smooth and the surface roughness is lower than 15 nm. Again, this topography is typical for amorphous films.

Generally, the chemical structure of the non-stoichiometric amorphous silicon oxynitride thin films can be described by the random bonding model (RBM) [25]. A similar concept can be applied also to Mg-Si-O-N films. In this model, silicon atoms randomly bond with O and N atoms to form a homogenous O-Si-N network with possible $SiO_4$, $SiO_3N$, $SiO_2N_2$, $SiON_3$ and $SiN_3$ tetrahedral coordinates. The presently studied thin films have also high amount of Mg content and it is likely that Mg taking on the role of a network former, in present of N. This assumption is based on the model for bulk oxynitride glasses, in which Mg is four-fold coordinated and likely participating in the silicon tetrahedral network due its high cation field strength [12, 26].

*3.2 Film properties*

*Film durability:*

Generally, corrosion by water is similar to acid corrosion in that alkali is removed from the glass surface and at high temperatures, water corrosion can become significant. After the treatment (keeping in moisture atmosphere at 85 °C for three day), the samples were observed by the naked eye and light microscope and were not noticeably affected by the treatment. Furthermore, reexamining by ellipsometry technique one year later after the first examination shows similar values as observed previously. This confirms that the Mg-Si-O-N thin films deposited onto float glass surface are chemically inert.

*Film thickness and roughness:*

The analyzed values of the Mg-Si-O-N thin films are reported in Table II. The thicknesses and roughness of the deposited thin films on float glass substrates were determined by MMSE. The film thicknesses range from 156 nm to 545 nm. Fig. 4 shows the relationship between the film thickness and the Mg content. It can be seen that the film thickness increases linearly with Mg content for all three series. Series M3 has higher values of thickness as compared with the series M1 and M2. The Mg content increases linearly with the Mg target power, as expected. The roughness of the Mg-Si-O-N films varies between 2 nm to 16 nm and roughly increases with increasing, the Mg target power as can be seen in Fig. 5 as a function of Mg content. M1 series have lower values of roughness as compared with the M2 and M3 series.

*Film density and molar volume:*

The density and molar volume values are given in Table I. The densities of the films were found to have values varies between 2.99 g/cm$^3$ and 3.09 g/cm$^3$. Molar volume values were calculated accordingly to the equation given in Ref. [12]. The calculated molar volume values vary between 6.54 and 6.72 cm$^3$/mol. Both density and molar volume do not exhibit any systematic dependency on the Mg and or N content.

*Mechanical properties:*

The hardness and reduced elastic modulus values for the Mg-Si-O-N thin films are given in Table II. Hardness values vs the Mg content is plotted in Fig. 6. The hardness (H) values range from 15 GPa to 21 GPa and increases with increasing Mg content. The reduced elastic modulus as a function of Mg content is shown in Fig. 7. The reduced elastic modulus (Er) of the Mg-Si-O-N films varies between 120 GPa and 176 GPa and shows similar dependence on composition as observed for the hardness. The hardness and reduced elastic modulus values of the Mg-Si-O-N thin films deposited on the float glass substrate show higher values than pure $SiO_x$ (, H= 8.3 GPa and Er =72 GPa). The same holds for $SiO_xN_y$ (H=12.5–15.8 GPa and Er=126–160 GPa), and, $SiN_y$ (H=17.1 GPa and Er = 166 GPa,) reported by Liu et *al* [27]. Furthermore, the Mg-Si-O-N thin films have higher values of hardness and reduced elastic modulus than SiN (H =10 GPa, Er = 91GPa) and SiON (H= 16 GPa and Er = 120 GPa) thin films prepared during this study. Again, the Mg-Si-O-N thin films have higher values of hardness and elastic modulus than the bulk Mg-Si-O-N glasses. The increase of hardness and elastic modulus values with the nitrogen content is well known in both bulk and thin films. The addition of nitrogen increases cross-linking in the glass network mainly due to tri-coordinated nitrogen in the tetrahedral units. Previous investigations in the bulk and thin film Mg-Si-O-N systems [12, 14, 15, 28] show that the properties are also dependent on Mg content. For the present Mg-Si-O-N thin films both hardness and reduced elastic modulus display large dependence on the Mg content. The M2 and M3 series have almost similar values of hardness for the similar power of Mg target but the M3 series has higher values of reduced elastic modulus as compared to the M2 series. The M1 series has lower values of hardness and reduced elastic modulus. Generally, a high hardness of the amorphous material is accompanied by a high elastic modulus, with the exception of the sample ID# M-310 have comparatively low hardness value than the M-214 and M-314 samples, but have high reduced elastic modulus (176 GPa) as compare to M-214 (166 GPa) and M-314 (162 GPa).

*Optical properties:*

The values of refractive index ($n_r$) and extinction coefficients (k) for each sample and band gap (E) for M2 and M3 series are reported in Table II. The $n_r$, k and E values were obtained from Tauc-Lorentz model fitted to the ellipsometry data collected for each sample. $n_r$ and k in the Mg-Si-O-N system as a function of wavelength are shown in Fig. 8a and 8b respectively. As it can be observed, the shape of the spectra is similar for all samples. The refractive index at wavelength 633 nm varies between 1.89 and 2.00 and becomes higher for higher values of the Mg content as shown in Fig. 9.

The optical properties of the material not only depend on the chemical compositions but also correlated with surface and interface morphologies, defect structure and packing density. The refractive index increase with both Mg, predominantly, and N content for the present Mg-Si-O-N thin films. However, an attempt to ascertain these individual effects independently is hindered by the simultaneous increase of Mg and N contents. Silicon oxynitride thin films have intermediate refractive indices of between 1.50 and 2.00 and depend on the composition ratio of O/N in the films. As the Mg-Si-O-N thin films are nitrogen rich the $n_r$ values remain close to the pure $Si_3N_4$ (2.02) [29] . Thin films prepared in series M3 have slightly higher values of refractive index as compared to the series M1 and M2 prepared with similar Mg target power. This might be due to the high amount of Mg content in the series M3. In contrast to bulk oxynitride glasses, Mg-Si-O-N thin films have higher values of refractive index as compare to the Ca-Si-O-N thin films prepared by the similar deposition technique [11]. Previous investigation on bulk AE-Si-O-N glasses, where AE= alkaline earth metals, show that Mg containing glasses have lower values of refractive indices as compare to Ca, Sr and Ba [12]. Furthermore, Mg-Si-O-N thin films have higher $n_r$ values compare to the bulk Mg-Si-O-N glasses.

There are two possible reasons for this matter; first, the Mg-Si-O-N thin films contain about 2.0 times more N and 1.5 times more Mg than the bulk Mg-Si-O-N glasses prepared by traditional quench melting. This means that thin films can grow with less Si and O contents as compared to amorphous bulk materials. Therefore, a denser structure is obtained with high N content, due to the presence of three coordinated N as compared to the two coordinated O. Secondly, these thin films exhibit higher density values (2.98– 3.09 g/cm$^3$) and lower molar volume (6.54 and 6.75 cm$^3$/mol ) than the bulk Mg-Si-(Al)-O-N glasses having density values (2.55 – 2.86) [12, 15, 28] and molar volume values (7.03 – 7.23) [12]. In general, the $n_r$ is directly related to the materials density and molecular polarizability and increased with increasing density and polarizability of the material. Therefore, the high $n_r$ values of the Mg-Si-O-N thin films are due to the changes in the film density as well as polarizability, or a combined effect thereof. Furthermore, the density of these films is close to crystalline $Si_3N_4$ (3.1-3.3 g/cm$^3$) as compared to crystalline $SiO_2$ (2.2-2.6 g/cm$^3$) and amorphous $SiO_2$ (2.2 g/cm$^3$). Thin films show high refractive indices and high transparency is considered to be useful materials for optical applications.

The extinction coefficients at the wavelength 233 nm vary between 0.011 and 0.186 and are predominately affected by the Mg content as shown in Fig. 10. The extinction coefficient (k) of all films at 233 nm is less than 0.20 and was found to be strongly dependent on the Mg content with the exception of the sample with ID: M-314 [Table II]. This indicates that the Mg-Si-O-N thin films exhibit sufficiently high visible light transparency. The M3 series have higher values of k than the M1 and M2 series. The k values were found to be zero in the visible region as can be seen in Fig. 8b. However, the k amplitude significantly increases to become higher near ultraviolet region (lower wavelengths), due to a progressive replacement of Si(ON) and Mg(ON) bonds by highly absorbent Si-Si and Mg-Si bonds. The band gap value as a function of Mg content for series M2 and M3 is shown in Fig. 11. The band gap values vary between 3.436 and 4.490 and decrease with increasing Mg content.

*3.3 Influence of substrate temperature on properties.*

In order to determine the influence of the substrate temperature on the properties of the Mg-Si-O-N thin films, samples were deposited on float glass substrate at temperatures of 100, 350 and 510 °C (M4 series). Identical synthesis parameters, i.e. Mg target power of 70 W, Si target power of 50, Ar flow of 31.4 sccm, $N_2$ flow of 8 sccm, and $O_2$ flow of 0.6 sccm were used for all these depositions. The composition obtained by XPS is shown in table III and the properties are reported in table IV respectively. The films grown at different substrate temperatures were transparent, homogeneous and amorphous. Accordingly, to XPS data, the composition of the film depends on the substrate temperature. The content of Mg decrease and of N increases in the films that are prepared at increasing substrate temperatures. The thickness of the films decreased from 233 nm to 169 nm and the roughness decrease from 3.42 nm to 1.06 with increasing the substrate temperate from 100 °C to 510 °C. The density of the Mg-Si-O-N thin films varies from 2.98 to 3.05 g/cm$^3$ and increases with increasing substrate temperature, a similar trend was also reported by Patsalas et al [30] and Aarik et al [31] . The molar volume values vary from 6.80 to 6.65 cm$^3$/mol and decrease with increasing substrate temperature. The data indicate that increasing the N content decreases the molar volume and increasing the Mg content increases it. The present results agree with those of previous studies of bulk glasses in the alkaline-earth-Si-O-N systems [12, 32-33]. The thin film deposited at substrate temperature of 100 °C have lower value of hardness and higher values of reduced elastic modulus than the films deposited at 350 and 510 °C. The hardness and elastic modulus values of thin films deposited at 350 and 510 °C have almost similar values. The refractive index values are also affected by the substrate temperature and show a lower value of refractive index for the thin film deposited at a substrate temperature of 100 °C. A similar trend was also observed for $HfO_2$ thin films [31]. The extinction coefficient values decrease with increasing substrate temperature. These variations in properties are particularly important when considering temperature-sensitive substrates such as polymers. The thin film in the Mg-Si-O-N system deposit at 100 °C substrate temperature has a higher hardness (10 GPa) and reduced elastic modulus (124 GPa) than uncoated float glass and polymer substrates and thus can be used as protective coating for polymer substrates.

## 4. Conclusions

Amorphous Mg-Si-O-N thin films in a large range of composition were grown by reactive RF magnetron sputtering in an UHV deposition system. The Mg-Si-O-N thin films containing high amount of magnesium metal content up to 28 at.%. The composition of the thin film is mainly controlled by the power of the Mg target and the substrate temperature. The hardness and reduced elastic modulus were determined via nano-indentation technique and are found to be up to 21 GPa and 176 GPa respectively. The hardness and reduced elastic modulus values are higher than Si3N4 and SiON thin films and bulk Mg-Si-O-N glasses. These results indicate that Mg contributes to the high hardness and elastic moduli values of the Mg-Si-O-N thin films due to its high cationic field strength. The refractive index values range from 1.89 to 2.01 and are found to increase with the Mg (predominately) and N contents. The obtained thin films in the Mg-Si-O-N are transparent, having extinction coefficient values of zero at the visible region and band gap values $\geq$ 3.2 eV. The Mg-Si-O-N thin films of series M3 have high optical quality, better mechanical and chemical properties as compared with series M1 and M2. Furthermore, thin films deposit at 100 oC substrate temperature can also be as a protective coating for polymer substrates due to its higher hardness (10 GPa) and reduced elastic modulus (124 GPa).

## Acknowledgements

This work was supported by the Vinnova (Grant No. 2015-04809) and ÅForsk foundation (Grant No. 14- 457). We also acknowledge support from the European Research Council under the European Community's Seventh Framework Programme (FP/2007-2013) / ERC grant agreement no 335383 and the Swedish Foundation for Strategic Research (SSF) through the Future Research Leaders 5 Program (B.P. and P.E), the Swedish Government Strategic Research Area in Materials Science on Functional Materials at Linköping University ( P. E., J.B) Faculty Grant SFO Mat LiU NO 2009 00971. The authors acknowledge Grzegorz Greczynski for XPS analysis, Esteban Broitman for nanoindentation analysis, Jörgen Bengtsson for AFM imaging, and Ludvig Landälv for assistance with the sputtering equipment.

Table I. Film designation, Determined film compositions [atomic % defined as Mg/(Mg+Si), N/(N+O)] , Density (ρ), and Molar volume (MV). *Series M2 has been previously reported; here the values are given for comparison purpose.

| ID | Thin film composition (atomic %) | Mg/ (Mg+Si) | N/ (N+O) | ρ g/cm$^3$ | MV cm$^3$/mol |
|---|---|---|---|---|---|
| Series-M1, Ar = 31.4 sccm, $N_2$ = 8 sccm, $O_2$ = 0.6 sccm | | | | | |
| M-106 | $Mg_{17.0}Si_{28.0}O_{25.0}N_{30.0}$ | 37.78 | 54.55 | 3.09 | 6.54 |
| M-110 | $Mg_{19.0}Si_{26.0}O_{24.0}N_{32.0}$ | 42.22 | 57.14 | 3.02 | 6.75 |
| M-114 | $Mg_{21.0}Si_{27.0}O_{16.0}N_{37.0}$ | 43.75 | 69.81 | - | - |
| Series-M2*, Ar = 31.4 sccm, $N_2$ = 8 sccm, $O_2$ = 0.6 sccm | | | | | |
| M-206 | $Mg_{15.2}Si_{29.5}O_{19.5}N_{36.0}$ | 33.70 | 64.87 | 2.99 | 6.72 |
| M-210 | $Mg_{16.8}Si_{29.5}O_{15.5}N_{40.0}$ | 36.28 | 72.07 | 3.01 | 6.68 |
| M-214 | $Mg_{20.0}Si_{25.0}O_{24.0}N_{31.0}$ | 44.50 | 56.36 | 2.99 | 6.71 |
| Series-M3, Ar = 27.4 sccm, $N_2$ = 12 sccm, $O_2$ = 0.6 sccm | | | | | |
| M-306 | $Mg_{17.0}Si_{27.7}O_{21.0}N_{34.3}$ | 38.03 | 62.02 | - | - |
| M-310 | $Mg_{24.9}Si_{22.0}O_{21.0}N_{32.0}$ | 53.09 | 60.38 | 3.00 | 6.70 |
| M-314 | $Mg_{28.0}\ Si_{20.0}O_{22.0}N_{31.0}$ | 58.33 | 58.49 | 2.99 | 6.72 |

Table II. Film designation, Film thickness, Roughness, Hardness (H), Elastic modulus ($E_r$), Refractive index ($n_r$) Extinction coefficient (k) and Band gap (E) for films deposited on float glass. The estimated errors in H and $E_r$ do not exceed ± 2 GPa and ± 8 GPa, respectively.

| ID | Thickness (nm) | Roughness (nm) | H (GPa) | $E_r$ (GPa) | $n_r$ at 633 nm | k at 233 nm | E (eV) |
|---|---|---|---|---|---|---|---|
| $Si_3N_4$ | | | 10.1 | 91 | 2.07 | 0.017 | - |
| SiON | | | 15.5 | 120 | 1.65 | 0.019 | - |
| M-106 | 265 | 2.54 | 16.8 | 124 | 1.89 | 0.042 | - |
| M-110 | 285 | 2.92 | 17.3 | 126 | 1.92 | 0.056 | - |
| M-114 | 157 | 3.79 | 17.6 | 119 | 1.96 | 0.151 | - |
| M-206 | 418 | 6.93 | 18.1 | 151 | 1.90 | 0.033 | 4.31 |
| M-210 | 449 | 8.82 | 17.9 | 131 | 1.95 | 0.071 | 4.02 |
| M-214 | 463 | 9.35 | 20.6 | 166 | 1.99 | 0.012 | 3.73 |
| M-306 | 365 | 7.80 | 18.2 | 151 | 1.90 | 0.051 | 4.02 |
| M-310 | 481 | 16.0 | 20.1 | 176 | 1.96 | 0.186 | 3.44 |
| M-314 | 545 | 16.6 | 20.4 | 162 | 2.01 | 0.160 | 3.09 |

Table III. Film designation, Determined film compositions (atomic %), Substrate temperature (°C) Mg/(Mg+Si), N/(N+O) , Density (ρ) and Molar volume (MV).

| ID | Thin film composition (atomic %) | Substrate temp (°C) | Mg/ (Mg+Si) | N/ (N+O) | ρ g/cm$^3$ | MV cm$^3$/mol |
|---|---|---|---|---|---|---|
| Series-M4, Ar = 31.4 sccm, $N_2$ = 8 sccm, $O_2$ = 0.6 sccm | | | | | | |
| M-10 | $Mg_{30.0}Si_{16.0}O_{45.0}N_{9.0}$ | 100 | 65.22 | 16.67 | 2.98 | 6.80 |
| M-35 | $Mg_{25.0}Si_{21.0}O_{32.0}N_{22.0}$ | 350 | 54.35 | 40.74 | 2.99 | 6.75 |
| M-51 | $Mg_{12.0}Si_{31.0}O_{33.0}N_{24.0}$ | 510 | 27.27 | 42.11 | 3.05 | 6.65 |

Table IV. Film designation, Film thickness, Roughness, Hardness (H), Elastic modulus ($E_r$), Refractive index ($n_r$) and Extinction coefficient (k) for films deposited on float glass substrate. The estimated errors in H and $E_r$ do not exceed ± 1 GPa and ± 3 GPa, respectively.

| ID | Thickness (nm) | Roughness (nm) | H (GPa) | $E_r$ (GPa) | $n_r$ at 633 nm | k at 233 nm |
|---|---|---|---|---|---|---|
| M-10 | 232.73 | 3.42 | 10.1 | 124 | 1.75 | 0.002 |
| M-35 | 133.48 | 2.45 | 14.6 | 122 | 1.84 | 0.001 |
| M-51 | 168.45 | 1.06 | 14.6 | 115 | 1.81 | 0.000 |

Figure Captions:

*Fig.1: Mg content as a function of Mg target power for Mg-Si-O-N films*

*Fig. 2: (a). Optical image of Mg-Si-O-N thin film ( ID: M-210 ) on float glass, silicon wafer and sapphire substrates, prepared with Mg target power 100 W, Si target power 100 W and N2 flow 30%. (b) Bulk Mg-Si-O-N glass prepared by quench melting*

*Fig. 3: (a) SEM cross section image, (b) SEM back-scattered electron image and (c) AFM image of* $Mg_{24.9}Si_{22.0}O_{21.0}N_{32.0}$ *film*

*Fig. 4: Variation of thickness vs Mg content for Mg-Si-O-N films.*

*Fig. 5: Variation of roughness vs Mg content for Mg-Si-O-N films.*

*Fig. 6: Variation of hardness vs Mg content for Mg-Si-O-N films. Error bars correspond to the standard deviation*

*Fig. 7: Variation of reduced elastic modulus vs Mg content for Mg-Si-O-N films. Error bars correspond to the standard deviation*

*Fig. 8a: Variation of refractive index as a function of wavelength for Mg-Si-O-N films*

*Fig. 8b: Variation of extinction coefficient as a function of wavelength for Mg-Si-O-N films*

*Fig. 9: Variation of refractive index vs Mg content for Mg-Si-O-N films.*

*Fig. 10: Variation of extinction coefficient vs Mg content for Mg-Si-O-N films.*

*Fig. 11: Variation of Bandgap vs Mg content for Mg-Si-O-N films.*

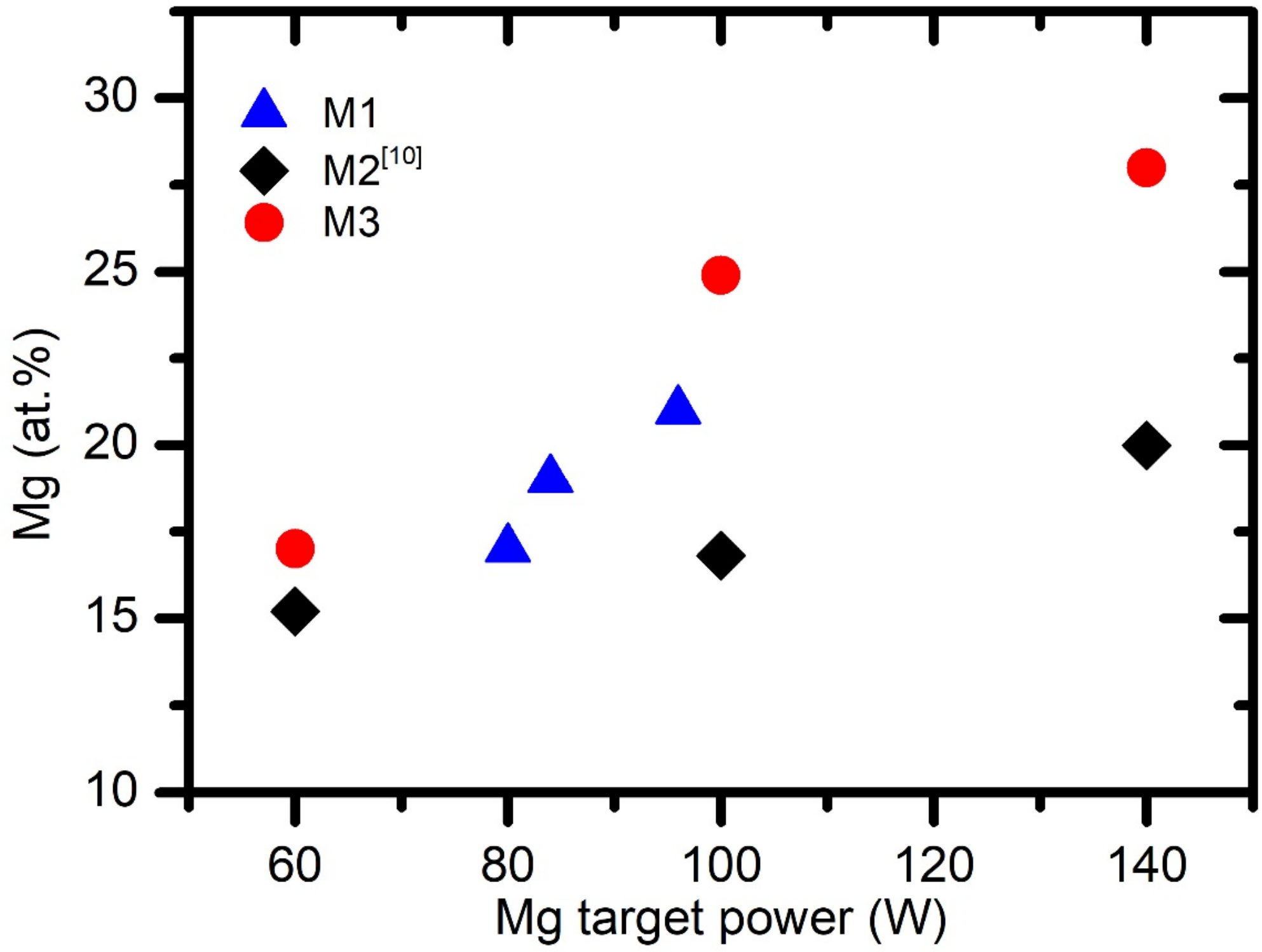

*Fig.1*

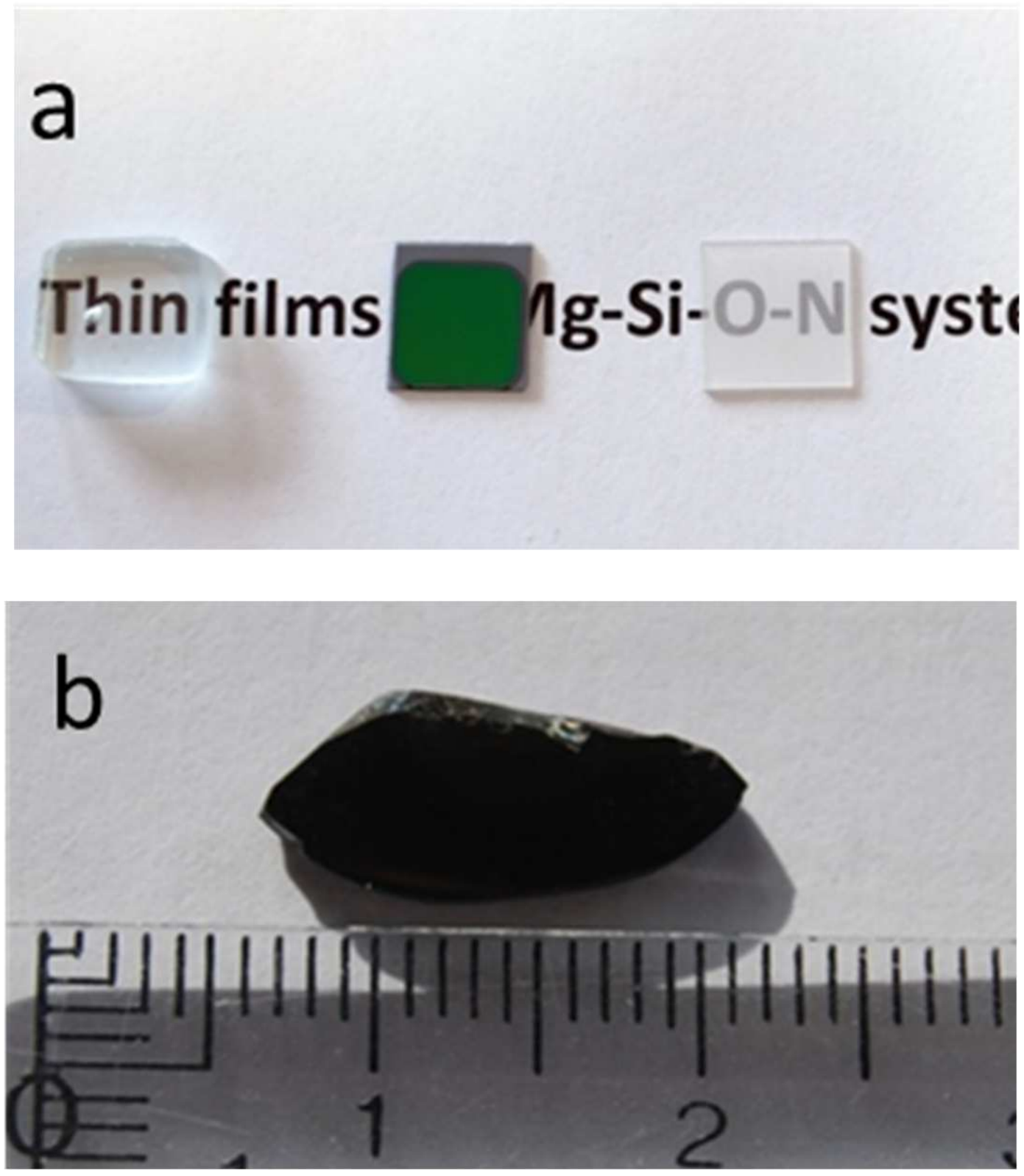

*Fig.2*

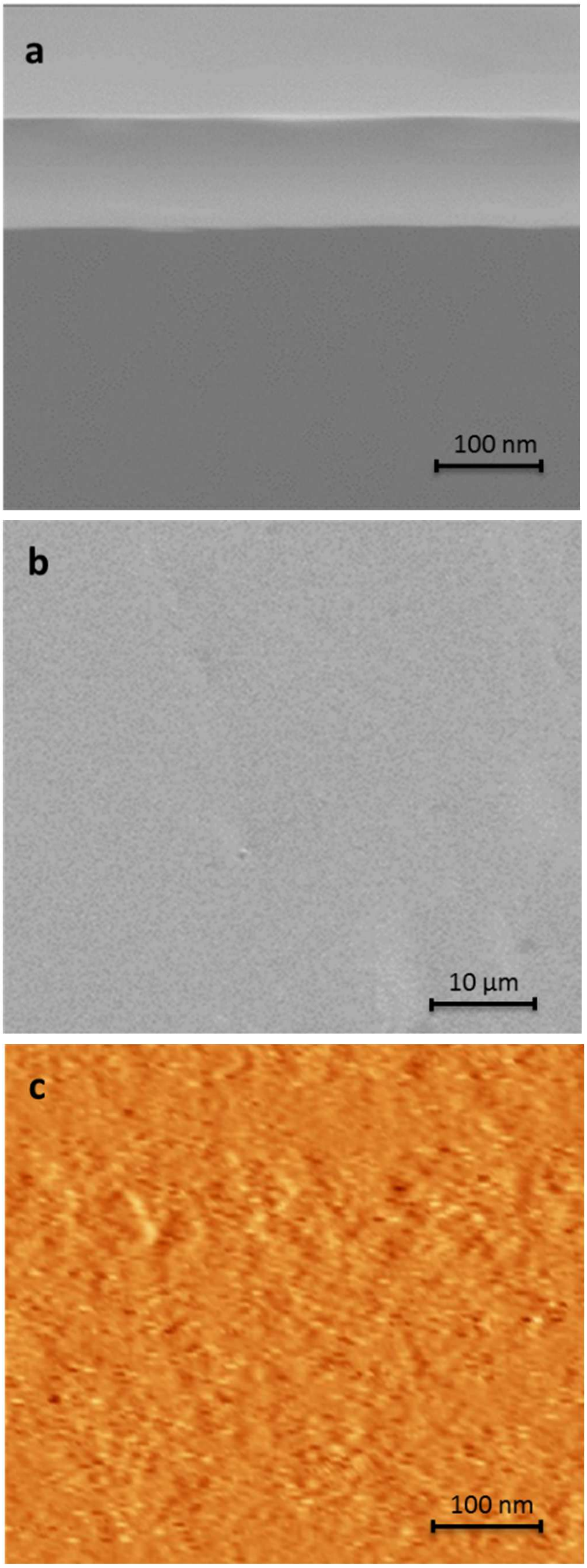

*Fig.3*

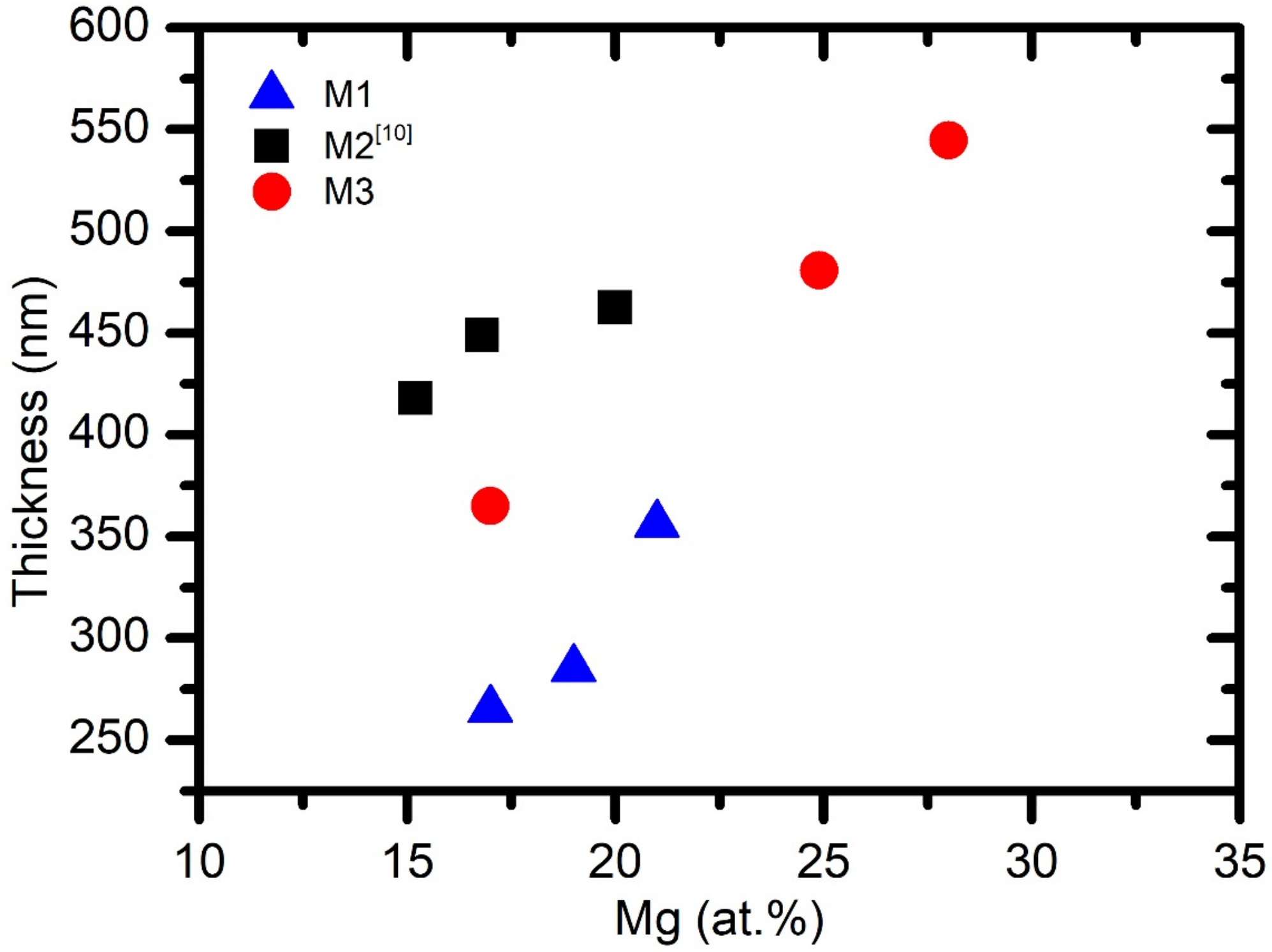

*Fig.4*

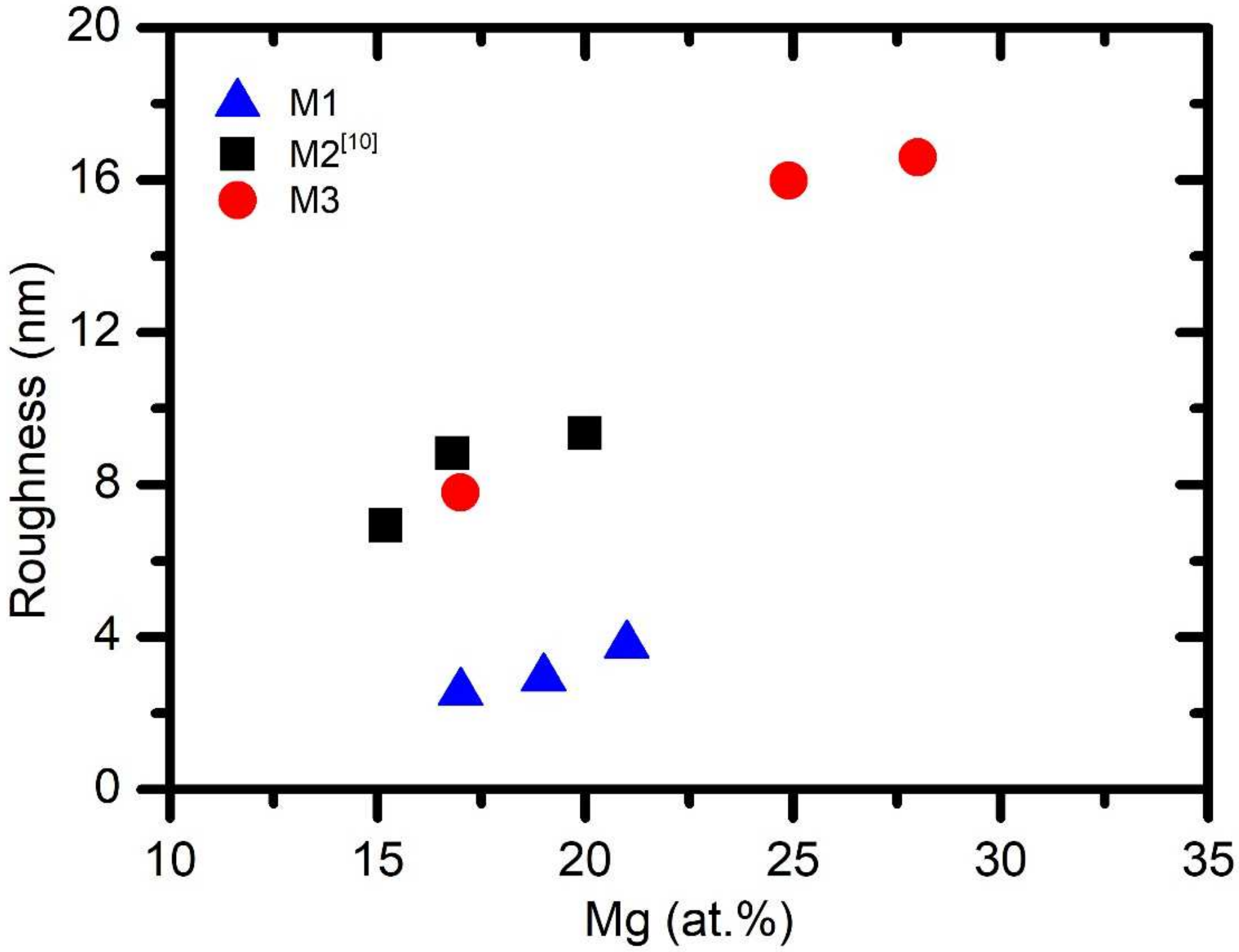

*Fig.5*

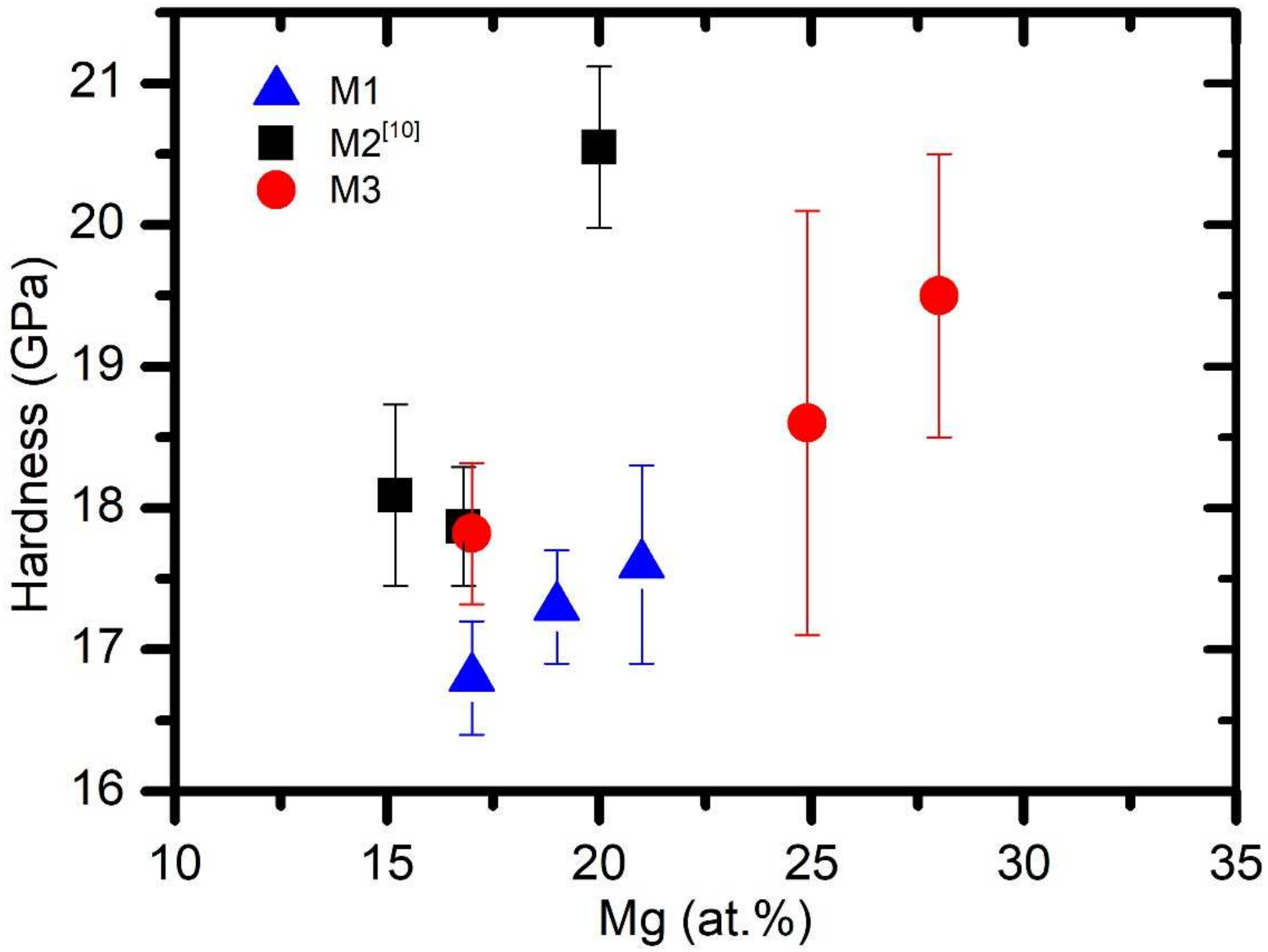

*Fig.6*

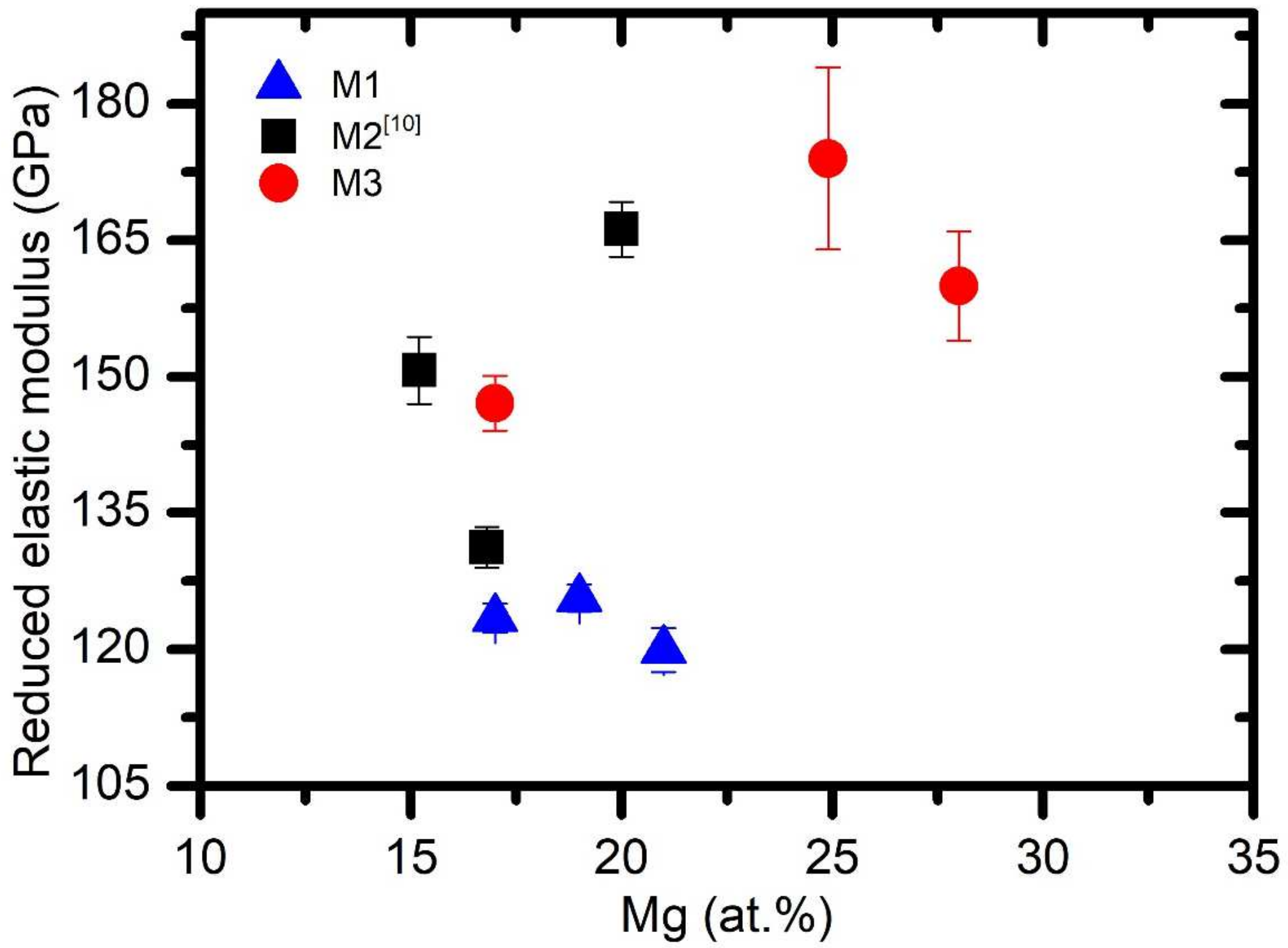

*Fig.7*

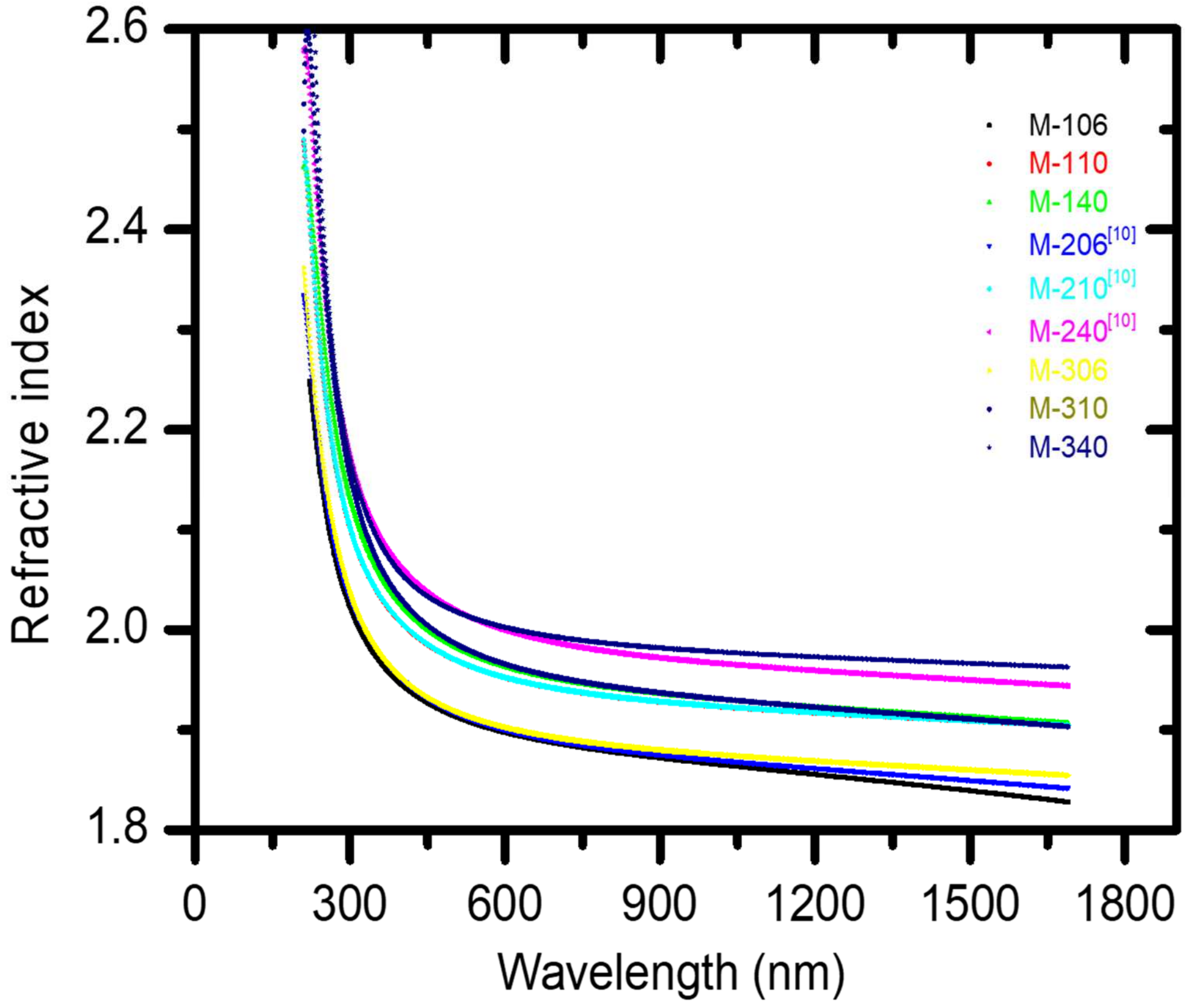

*Fig.8a*

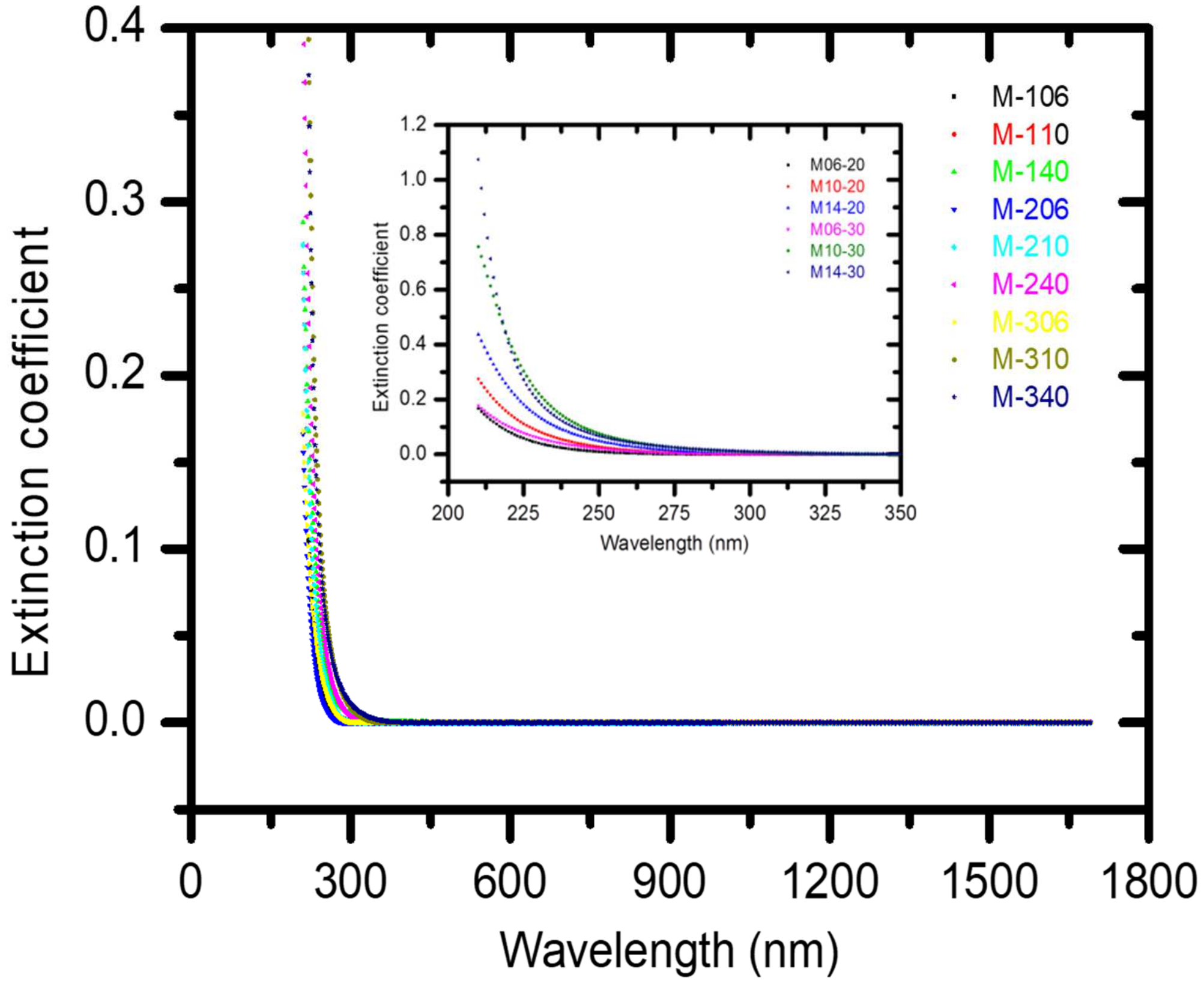

*Fig.8b*

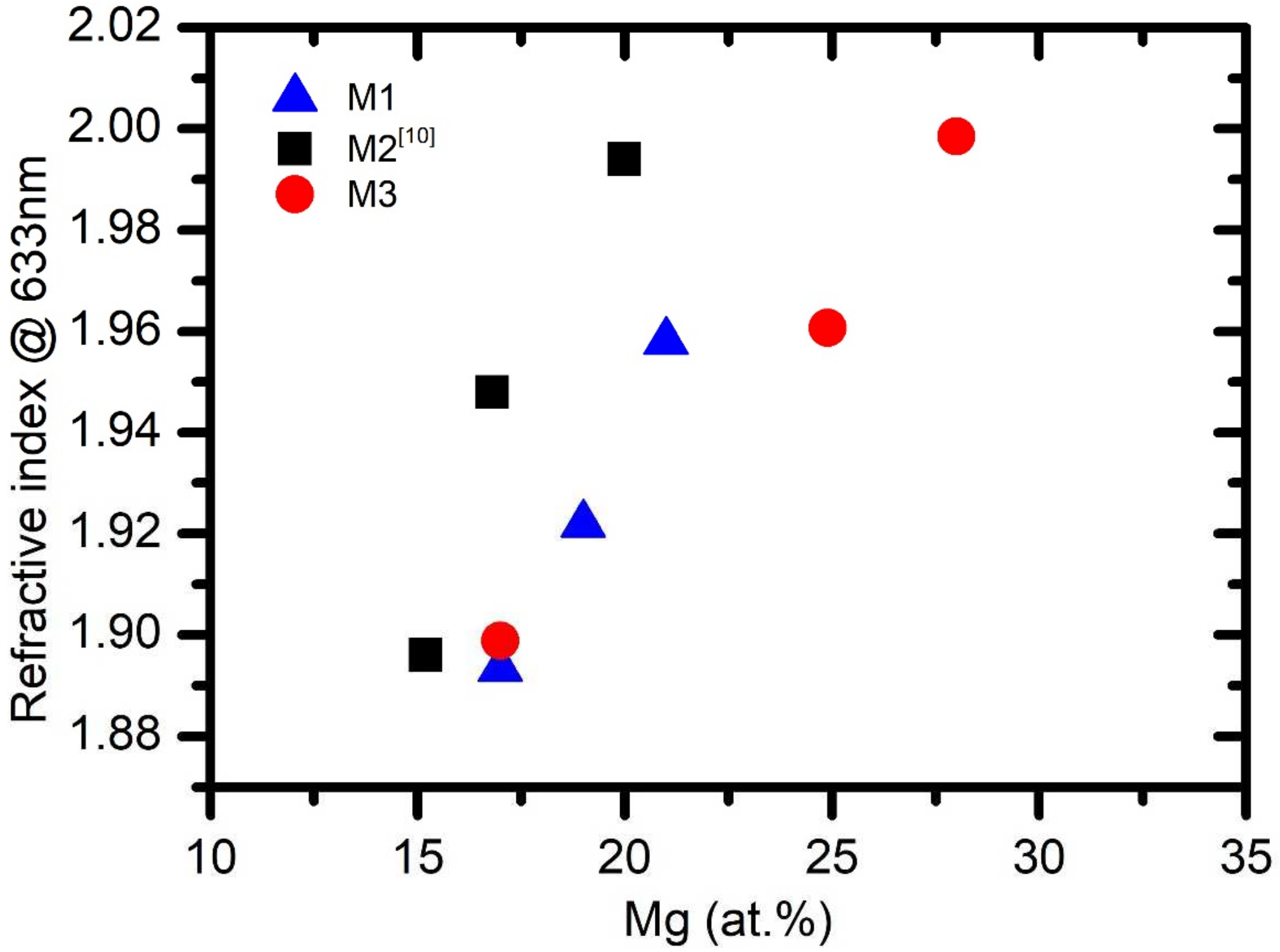

*Fig.9*

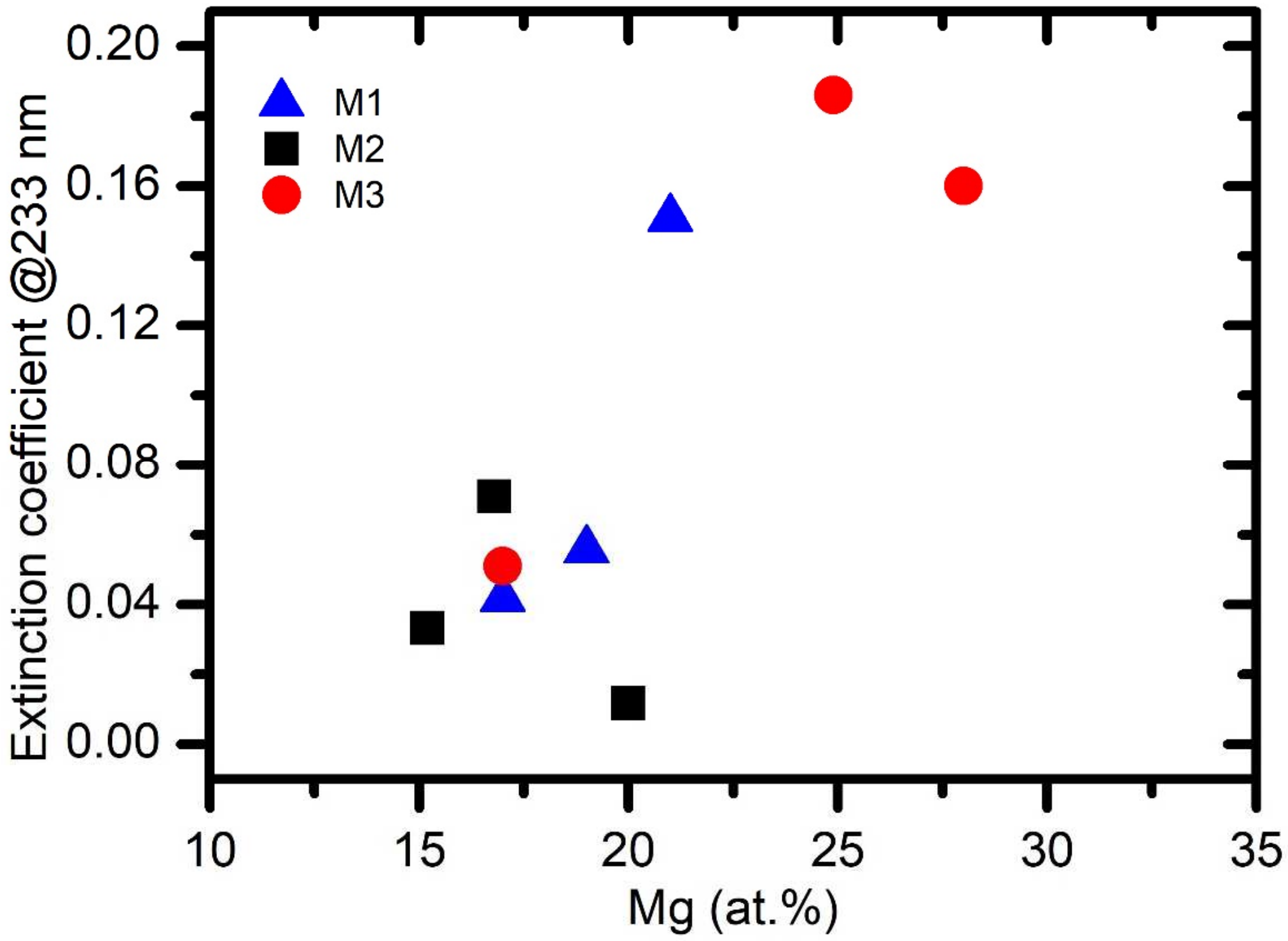

*Fig.10*

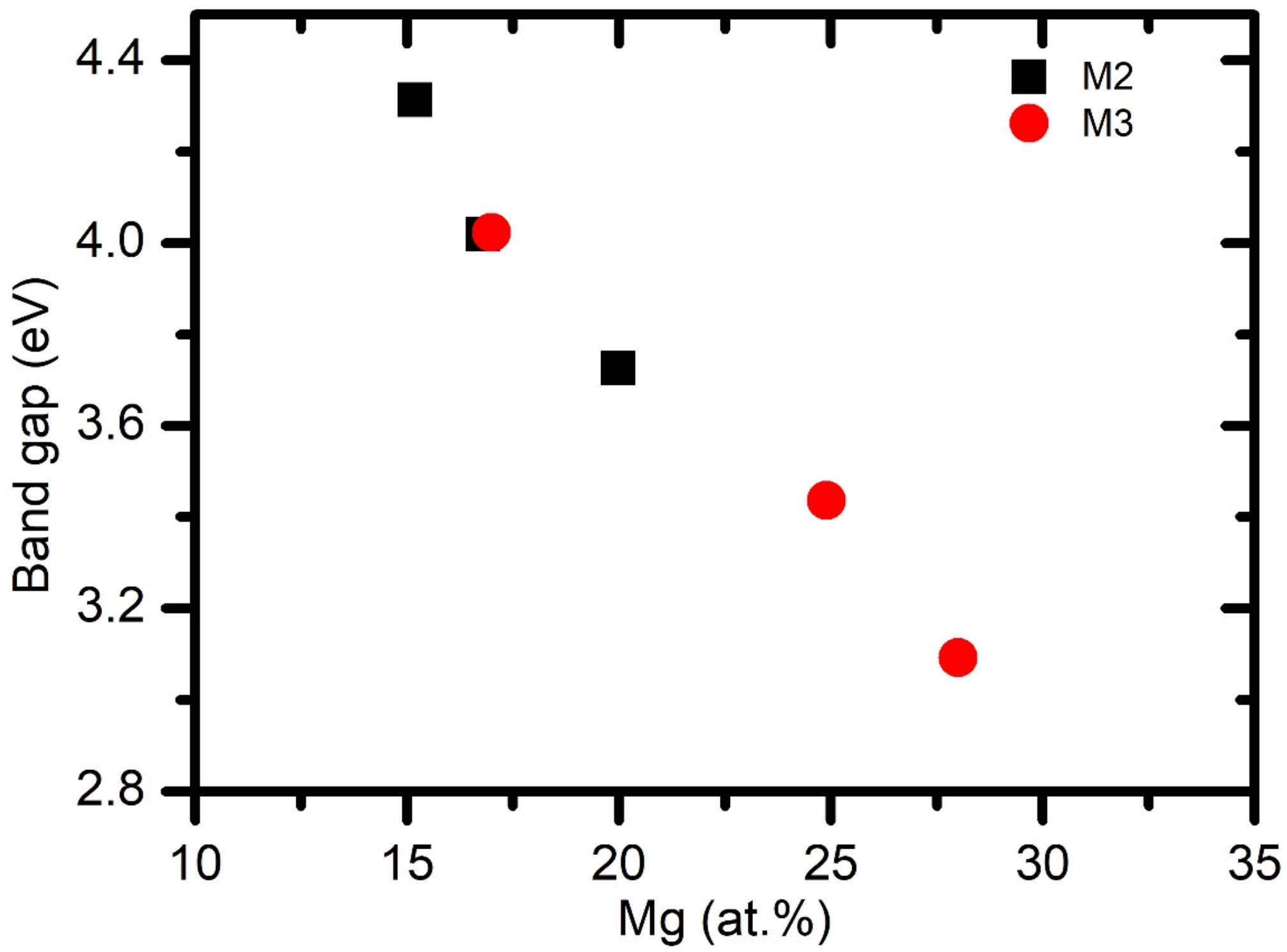

*Fig.11*